\documentclass[prd, twocolumn, nofootinbib, floatfix]{revtex4-1}

\usepackage{amsmath}
\usepackage{graphicx}
\usepackage{dcolumn}
\usepackage{bm}
\usepackage{epsfig}
\usepackage{amssymb,latexsym,mathrsfs}
\usepackage{graphicx}
\usepackage{color}
\usepackage{hyperref}

\hypersetup{
    colorlinks=true,
    linkcolor=red,
    citecolor=blue,
} 

\newcommand{\be}{\begin{equation}}
\newcommand{\ee}{\end{equation}}
\newcommand{\bs}{\begin{split}} 
\newcommand{\bea}{\begin{eqnarray}}
\newcommand{\eea}{\end{eqnarray}}

\newcommand{\om}{\Omega_m}

\newcommand{\ode}{\Omega_{\rm de}}

\newcommand{\lcdm}{$\Lambda$CDM} 
\newcommand{\gm}{G_{\rm matter}} 
 
\newcommand{\gl}{G_{\rm light}} 
\newcommand{\geff}{G_{\rm eff}}

\newcommand{\al}{\alpha} 
\newcommand{\alb}{\alpha_B} 
\newcommand{\albp}{\alpha_B'}

\newcommand{\fs}{f\sigma_8}

\begin{document}

\title{No Run Gravity} 

\author{Eric V.\ Linder${}^{1,2}$} 
\affiliation{${}^1$Berkeley Center for Cosmological Physics \& Berkeley Lab, 
University of California, Berkeley, CA 94720, USA\\ 
${}^2$Energetic Cosmos Laboratory, Nazarbayev University, 
Astana, Kazakhstan 010000}

\begin{abstract} 
Considering the dark energy/gravity landscape if next generation surveys 
of galaxies, cosmic microwave background radiation, and gravitational waves 
do not find clear modification of gravity, we develop No Run Gravity as a 
counterexample to the conclusion that this would imply general relativity 
with an expansion history described by an equation of state $w(z)$. 
No Run Gravity is a cubic Horndeski theory with a constant Planck mass, 
no gravitational slip, and no modification of gravitational waves, but a 
rich phenomenology beyond $w(z)$. We calculate the evolution of 
gravitational strength, sound speed, and cosmic growth within the 
theory and project sensitivities for upcoming DESI redshift space 
distortion data. 
\end{abstract}

\date{\today} 

\maketitle

\section{Introduction} 

Many signatures can appear in cosmic surveys to provide evidence of 
gravitational properties beyond general relativity. These can include 
modification of the tensor sector -- gravitational waves (GW) -- surveyed 
by laser interferometers such as LIGO/Virgo and LISA and by cosmic 
microwave background (CMB) B-mode polarization experiments such as 
Simons Observatory, CMB-S4, and LiteBIRD (see, e.g., \cite{1901.03321,so}. 
Already, deviation of GW propagation from the speed of light is effectively 
ruled out, at least in certain regimes \cite{gwspeed}. Damping of the GW 
amplitude over the time of 
propagation is allowed and related to the running of the Planck mass 
\cite{damp1,damp2,damp3}. 
Testing this through comparison of GW and electromagnetic luminosity 
distances to the same source will be highly interesting, especially since 
in certain theories it can be directly connected to suppression in the 
growth of cosmic structure \cite{nsg}. 

In the scalar -- density perturbation -- sector, signatures include 
modification of the growth of large scale structure, modification of the 
propagation of light (lensing), and gravitational slip, where the metric 
gravitational potentials differ from each other, unlike in general relativity. 
These can be probed by large scale structure surveys, through galaxy 
clustering and weak gravitational lensing, such as with DESI, Euclid, LSST, 
and WFIRST, and CMB surveys (see, e.g., \cite{reviewish,reviewfer} for 
recent reviews). 

This rich array of observational effects, and the phenomenology that goes 
with them, is comforting, as giving rise to hope that the excellent data 
will reveal key clues to the nature of cosmic acceleration and gravity. 
Suppose, however, that the data does not show significant deviation from 
general relativity -- no change in GW speed or damping, no gravitational 
slip. Change in the expansion history or growth history can be accommodated 
within general relativity by modification of the effective dark energy 
equation of state $w(z)$ from the concordance value $w=-1$ of the 
cosmological constant. Would such observations then imply the universe is 
described solely by general relativity with some $w(z)$? 

Such a question has been partially addressed by the introduction of 
No Slip Gravity \cite{nsg}, where a modified gravity theory was defined 
that had no change in GW speed or gravitational slip. This does have 
damping of GW amplitudes, however, as it involves a running Planck mass. 
Here we go one step further and consider no slip plus no running: No 
Run Gravity. This 
will give no change in the tensor sector (and so indeed may not be 
considered true modified gravity), and the two metric potentials will be 
the same. 

However -- the metric potentials do not have to be equal to 
Newton's constant, and so there can remain effects on cosmic growth and 
on light deflection in a manner distinct from any $w(z)$ (and indeed 
simple scalar field generalizations like k-essence \cite{kess1,kess2}). 
Thus we will 
continue to refer to it as a modified gravity theory in an informal way. 

No Run Gravity can be viewed as a minimal modification to general relativity. 
The hope is that it will serve as a benchmark to test deviations from 
general relativity and show the science reach of next generation surveys 
even when the more dramatic signatures may not be found. 

In Sec.~\ref{sec:method} we set up the theory and relate it to Horndeski 
gravity and effective field theory. Section~\ref{sec:evo} discusses 
the implications of simple parametrizations, including stability and early 
and late time limits. We project constraints on the theory from future 
data in Sec.~\ref{sec:fisher} and summarize and conclude in 
Sec.~\ref{sec:concl}.

\section{Theory of No Run Gravity} \label{sec:method} 

One can approach modified gravity theories through an effective field 
theory or property function approach, or the Horndeski Lagrangian for 
the most general scalar-tensor theory with second order equations of 
motion. The equivalence between these is given in detail in \cite{belsaw}. 
The property function approach has the advantage of being able to state 
the physical conditions succinctly: for GW speed to equal the speed of 
light then the property function $\al_T=0$, and for no running of the 
Planck mass then the property function $\al_M=0$. In fact, we want the 
more physical condition of vanishing slip. No Slip Gravity achieves this 
through $\al_B=-2\al_M$, but an appendix in 
\cite{nsg} (also see \cite{1711.04825}), briefly discussed an alternate 
method of accomplishing this by setting $\al_M=0$. We therefore define 
No Run Gravity as $\al_M=0=\al_T$. 

This leaves the braiding property function $\al_B$, and the observationally 
mostly moot kineticity $\al_K$ \cite{belsaw,nsgcmb}, as well as the 
expansion history in terms 
of the Hubble parameter $H(z)$ or effective dark energy equation of state 
$w(z)$. In terms of the Horndeski Lagrangian the theory is 
\be 
{\mathcal L}=\frac{1}{2}R+K(\phi,X)-G(\phi,X)\,\Box\phi\ , 
\ee 
where in the cosmic background $X=(1/2)(\dot\phi)^2$. This relates to 
the full Horndeski Lagrangian by setting $G_5=0$, $G_4=1/2$, and 
$G_3=G(\phi,x)$. 
Such a Lagrangian has been used in kinetic gravity braiding dark energy 
\cite{braid1}, inflation \cite{braid2}, and to solve the original 
cosmological constant problem \cite{temper}. 

Following \cite{belsaw} we can write the property functions as 
\bea 
\al_B&=&\frac{2\dot\phi X}{H}\,G_X\\ 
\al_K&=&\frac{12\dot\phi X}{H}\,(G_X+XG_{XX}) \\ 
&\qquad&+\frac{2X}{H^2}\,(K_X+2XK_{XX} 
-2G_\phi-2XG_{\phi X})\ , \notag 
\eea 
where subscripts $\phi$ and $X$ denote derivatives with respect to that 
variable. 
Note that a k-essence scalar field model, while having a sound speed 
degree of freedom, achieves this through the $K$ function; it has no 
$G$ function and so its $\al_B=0$. 

In the early universe, when $H\gg1$ (normalizing it by its value today), 
we might expect $\al_K\propto\al_B$. Similarly one can show that the 
effective dark energy density $\rho_{\rm de}\sim H\dot\phi XG_X$ under 
these circumstances so $\al_i\propto\ode(a)$, where 
$\ode(a)\sim\rho_{\rm de}/H^2$ is the dark energy density in 
units of the critical density. As very strongly cautioned by 
\cite{1512.06180} this relies on several assumptions, with the key ones 
being that a single Lagrangian function, e.g.\ $G_X$, dominates and that 
$H\gg1$; note \cite{1512.06180} emphasizes these hold for at best $z>10$ 
unless there is fine tuning, so well outside the range of most observational 
data. One can show that the theory can be ghost free and stable if 
$G_X\sim X^{n\ge-1}$. 

We choose to work with a shift symmetric theory, so we take $K_\phi=0$, 
$G_\phi=0$. In general we will take the background expansion to be given 
by $\Lambda$CDM.

\section{Gravity and Growth Evolution} \label{sec:evo} 

As mentioned, while No Run Gravity does not alter the tensor sector, 
and (by construction) has no difference between the two (time-time 
and space-space) metric potentials, hence no gravitational slip, it 
does have a modification in the {\it strength\/} of gravity. This is 
conventionally written in terms of the modified Poisson equations for 
nonrelativistic (matter) and relativistic (lightlike) particles, and 
referred to as $\gm$ and $\gl$. 

No Run Gravity has 
\bea 
\geff&\equiv&\gm=\gl=\frac{\alb+\albp}{\alb(1-\alb/2)+\albp} 
\label{eq:geffa}\\ 
&=&1+\frac{\alb^2}{\alb(2-\alb)+2\albp}\ , 
\label{eq:geffa2} 
\eea  
where we have normalized by Newton's constant so general relativity has 
$\gm=\gl=1$. A prime denotes $d/d\ln a$, where $a$ is the cosmic scale 
factor. We can rewrite this as a differential equation for $\alb$: 
\be 
\albp=\alb\,\left[-1+\frac{\alb}{2}\frac{\geff}{\geff-1}\right]\ . 
\label{eq:difa} 
\ee 

Thus we can either specify $\geff(a)$ and solve for $\alb$ from 
Eq.~(\ref{eq:difa}), or specify $\alb(a)$ and determine $\geff$ from 
Eq.~(\ref{eq:geffa}). To satisfy early universe constraints that gravity 
should look like general relativity for primordial nucleosynthesis and 
CMB last scattering, we want both $\alb$ and $\geff-1$ to vanish at 
early times, $a\ll1$. Suppose they evolve together 
such that $\alb/(\geff-1)=k$, with $k$ constant. This, together with 
$\alb\ll1$ at these early times, implies 
$\alb\propto(\geff-1)\propto a^{-1+k/2}$. If we wanted $\alb\propto\ode(a)$ 
at early times, this implies (for a \lcdm\ background) $k=8$. We emphasize 
though that this is only reasonable for very early times. 

The form for $\al_B$ can be quite varied, within stability considerations. 
One could also prefer to choose $\geff(a)$ instead and derive $\al_B$. In 
either case, we expect general relativity at early times ($\geff=1$, 
$\al_B=0$) and a frozen, constant value in a de Sitter future. The simplest, 
and most tractable, reasonable form would be the e-fold / $1+\tanh$ form 
used in \cite{nsg}. In this case, 
\be 
\alb(a)=\frac{A}{1+e^{-(\ln a-\ln a_t)/\tau}}=\frac{A}{1+(a/a_t)^{-1/\tau}}\ , 
\ee 
where $A$ is the amplitude of the transition 
(and maximum value of $\alb$; recall 
from \cite{nsg} that stability requires $\alb\ge0$), 
$a_t$ is the location of the transition, and $\tau$ is the sharpness of the 
transition in e-folds. 

An especially nice aspect of this form is that it corresponds to the identical 
functional form in $\geff-1$. That is, 
\be 
\geff(a)=1+\frac{A_G}{1+(a/a_{t,G})^{-1/\tau}}\ , 
\ee 
where 
\bea 
A_G&=&\frac{A}{2-A} \ , \label{eq:ag}\\ 
a_{t,G}&=&a_t\,\left(\frac{1+1/\tau}{1-A/2}\right)^\tau\ . \label{eq:atg} 
\eea 

If one forced $\alb\propto\ode(a)$ even beyond early times (which has 
no physical justification), this can be handled by this form with 
$\tau=1/3$ and $a_t=[\om/(1-\om)]^{1/3}\approx 0.75$, where $\om$ is 
the present dimensionless matter density $\om=1-\ode(a=1)$, which is a rather 
late transition (i.e.\ $\alb$ is halfway through its transition at 
$z=0.33$, and $\geff$ not until $z=-0.24$ in the future). 
However, the e-fold form is considerably more flexible. 

From Eq.~(\ref{eq:difa}) we can see that arbitrary assumed functional forms 
can easily fail. Looking at the term $\alb/(\geff-1)$, we see that if $\alb$ 
doesn't ``keep up'' with $\geff-1$ (even due to numerical noise in the 
nonlinear equation) then the other, $-1$ term is likely to 
dominate and $\albp$ will be driven negative. This implies that $\alb$ 
retreats to zero, and by Eq.~(\ref{eq:geffa}), $\geff$ is driven to one. 
That is, deviations from general relativity will not be described 
successfully. Conversely, if $\alb$ grows too fast then $\albp$ gets 
large and there is a runaway process that violates the stability condition. 
Note that we do not want $\alb\propto (\geff-1)$ in general; 
this too can runaway. 
Rather we want a balance in the evolution, such as demonstrated by the 
e-fold form's property that $\geff-1$ is a time lagged, scaled version 
of $\alb(a)$. 

\begin{figure}[htbp!]
\includegraphics[width=\columnwidth]{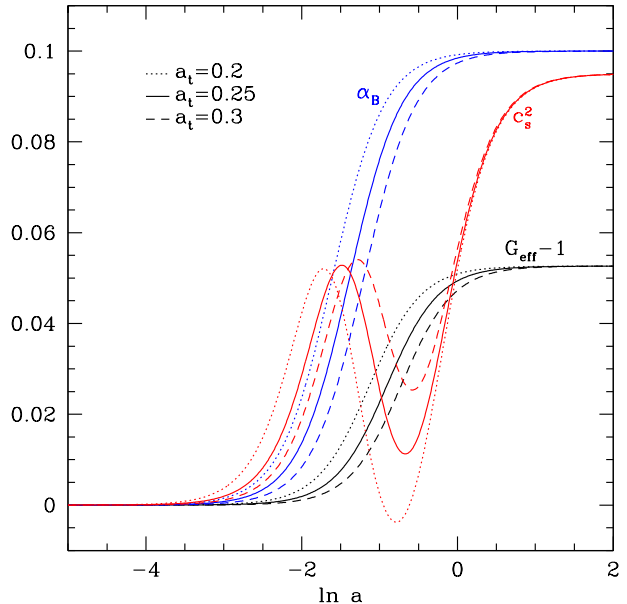} 
\caption{
An e-fold transition in the braiding property function $\alb$ (blue curves) 
induces a lagged, scaled e-fold modification of the gravitational strength 
$\geff$ (black curves) 
away from its general relativity value of one. Here the results are shown 
for three different values of the transition scale factor $a_t$, for fixed 
$A=0.1$, $\tau=1/3$. Too early a transition leads to the sound speed squared 
(red curves) going negative and an unstable model. 
} 
\label{fig:geffabtvary} 
\end{figure}

Figure~\ref{fig:geffabtvary} shows the delay and scaling between $\alb$ 
and $\geff-1$ for three values of $a_t$. Note that $\geff-1$ obeys these 
relations given by Eqs.~(\ref{eq:ag}) and (\ref{eq:atg}). Thus a transition 
in $\alb$ at $a_t=0.25$ does not give a transition in $\geff$ until 
$a_{t,G}=0.44$, and the maximum modification of the gravitational strength is 
slightly over half the maximum of $\alb$.

In addition to ensuring that deviation in the gravitational strength from 
general relativity can occur, and gravity remains positive, we also must 
ensure the stability of the model. This corresponds to the sound speed 
squared being nonnegative, $c_s^2\ge0$, and we plot this as well in 
Fig.~\ref{fig:geffabtvary}. 
In No Run Gravity, 
\be 
\al c_s^2=\alb\left(1-\frac{\alb}{2}\right)+\albp-\frac{3}{2}\om(a)\,\alb\ , 
\label{eq:cs2} 
\ee 
in the matter dominated era and later, where $\al=\al_K+(3/2)\alb^2\ge0$, 
and $\om(a)=1-\ode(a)$ is the dimensionless matter density. 

The condition $c_s^2\ge0$ is satisfied at early times if 
$\albp\ge\alb/2$, corresponding to $\tau\le2$, i.e.\ the transition takes 
less than two e-folds. In fact, since this is an early time limit we can 
state this more generally than an assumption of a steplike function over 
a long interval. Simply put, if at early times $\alb\sim a^{1/\tau}$ then 
we require $\tau\le2$. That is, $\alb$ must grow faster than $a^{1/2}$. 
In the late time limit, we require $\alb\le2$, or in the e-fold case, 
$A\le2$. 

Stability could be satisfied in both limits, but fail at some intermediate 
redshift, as seen in Fig.~\ref{fig:geffabtvary}. Generally one would have 
to scan numerically through the model parameter space, for all redshifts, 
to check the model is healthy. However, again a virtue of the e-fold model 
is that this can be done analytically for $\tau=1/3$, corresponding to 
early time deviations $\alb\sim a^3$. The solution to the boundary where 
$c_s^2=0$ is given by 
\be 
a_t^3=\frac{1}{32}\frac{\om}{1-\om}\left[14-A-\sqrt{15(2-A)(6-A)}\right]\ .
\ee 

\begin{figure}[htbp!]
\includegraphics[width=\columnwidth]{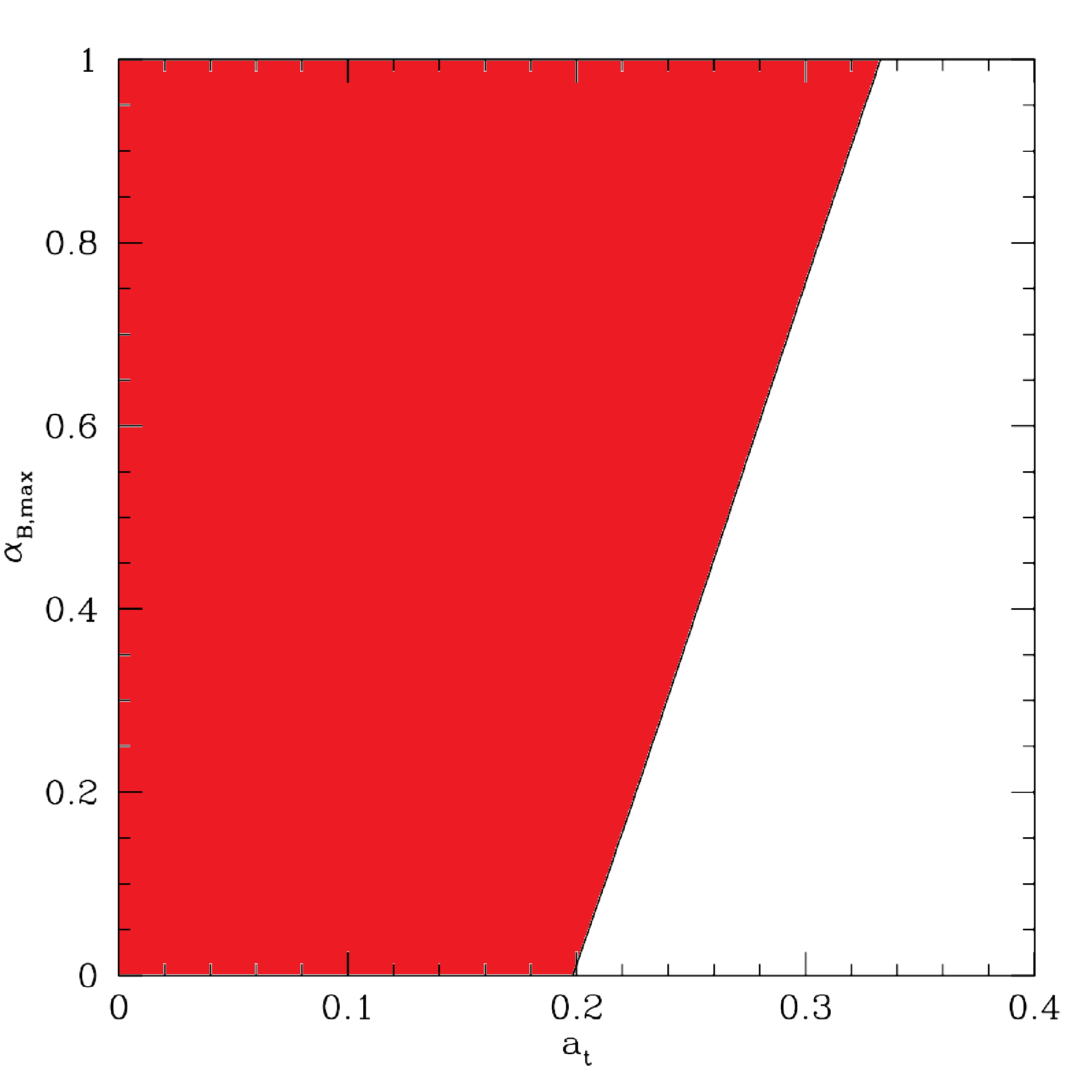} 
\caption{
The e-fold model parameter space of the maximum amplitude 
$A\equiv\al_{B,{\rm max}}$ and transition scale factor $a_t$ can be 
divided into an unstable region (red shaded) where at some redshift 
$c_s^2<0$ and a healthy region (unshaded). Here $\tau=1/3$. 
} 
\label{fig:stabshade} 
\end{figure}

Figure~\ref{fig:stabshade} illustrates the constraint in the $A$--$a_t$ 
plane, with the unhealthy region where at some redshift $c_s^2<0$ shaded 
red. Interestingly, a quite good approximation to the boundary is given 
by a simple linear function: $a_t=0.198+0.135\,A$. Apart from soundness, 
one could add observational limits, such as that gravity is no more than 
10\% stronger than general relativity. This would cut across the stable 
region of the figure at approximately $\al_{B,{\rm max}}\lesssim 0.2$. We 
consider observational constraints more rigorously in Sec.~\ref{sec:fisher}.

One could extend the e-fold form to multiple plateaus, either in 
$\geff$ or $\alb$, e.g.\ 
\be 
\alb=\frac{A_1}{1+(a/a_1)^{-1/\tau_1}}+\frac{A_2-A_1}{1+(a/a_2)^{-1/\tau_2}} 
\ , 
\ee 
and calculate $\geff$ analytically or numerically. This offers the 
freedom of having a further step up or step down after an intermediate 
phase. Indeed with $N$ steps one could approximate any desired function. 
However, the number of model parameters doubles with each step and anything 
beyond the simple case will be difficult to constrain observationally. 

\begin{figure}[htbp!]
\includegraphics[width=\columnwidth]{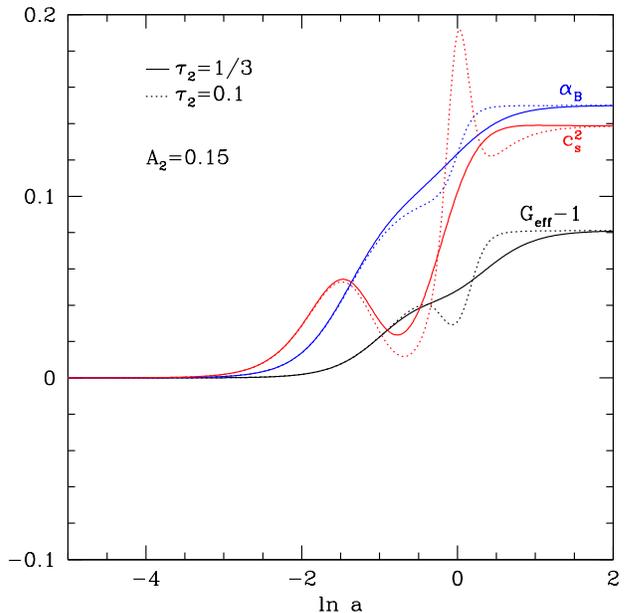}\\ 
\includegraphics[width=\columnwidth]{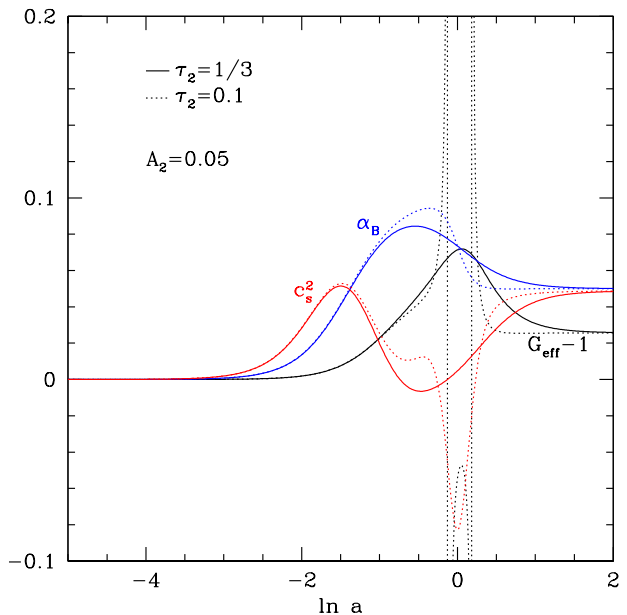} 
\caption{
For a more complicated dependence of $\alb(a)$ the behavior of $\geff(a)$ 
and $c_s^2(a)$ can be nonmonotonic, negative, or divergent. Here $\alb(a)$ 
is given by two e-fold transitions, with $A_1=0.1$, $a_1=0.25$, $\tau=1/3$ 
as before but $a_2=1$ and $\tau_2=1/3$ (solid curves) or 0.1 (dotted). The 
top panel shows a step up, with $A_2=A_1+0.05$; the bottom panel shows a 
step down, with $A_2=A_1-0.05$. 
} 
\label{fig:geff2} 
\end{figure}

Furthermore, a reduction in $\albp$, or even its negative value for 
a decreasing $\albp(a)$, can turn models unstable that were stable for 
the single e-fold form. Moreover, a monotonically increasing $\alb$ can 
still give rise to a nonmonotonic $\geff$ due to the interaction of 
$\alb$ and $\albp$ in Eq.~(\ref{eq:geffa}). We demonstrate these properties 
in Fig.~\ref{fig:geff2} for a second e-fold transition stepping either 
up or down.

Note that a period of decreasing $\alb$, i.e.\ negative $\albp$, can be 
dangerous, as $\geff$ diverges when $\albp=-\alb(1-\alb/2)$. A divergence 
necessarily means that $c_s^2<0$, as by Eq.~(\ref{eq:cs2}) the value of 
$c_s^2$ is the denominator of $\geff$ minus $(3/2)\om(a)\alb$. Conversely, 
one can have $c_s^2<0$ without $\geff$ being far from one. However, 
a decreasing $\alb$ can enhance $\geff$ without causing it to diverge. 

\begin{figure}[htbp!]
\includegraphics[width=\columnwidth]{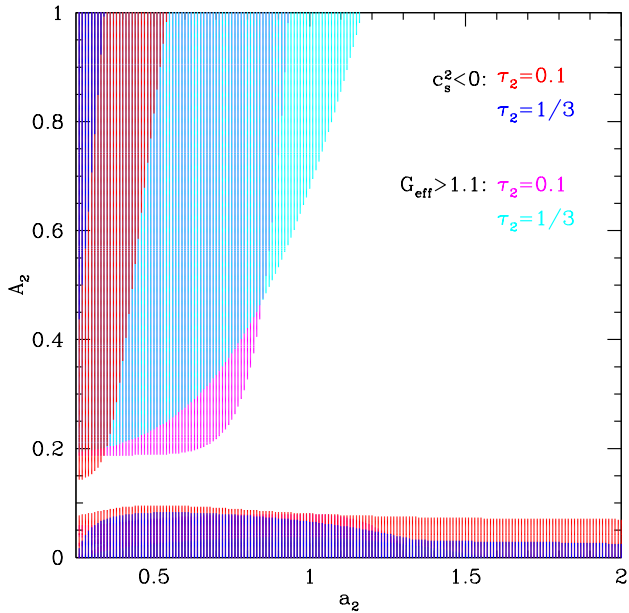}
\caption{
The instability and high gravity regions of the parameter space are shown 
for the two e-fold transition form of $\alb(a)$. The first transition has 
$A_1=0.1$, $a_1=0.25$, $\tau=1/3$ as before and the second has $\tau_2=1/3$ 
(red or magenta) or 0.1 (blue or cyan). Regions in the $a_2$--$A_2$ plane 
are shaded if they are unstable with $c_s^2<0$ (red/blue) or have high 
gravity, $\geff>1.1$ for $a\le1$ (magenta/cyan). For the lower shaded region 
the high gravity region lies wholly within the unstable region and is not 
shown. 
} 
\label{fig:stab2} 
\end{figure}

Figure~\ref{fig:stab2} shows the regions of instability and of high 
gravity ($\geff>1.1$ for $a\le1$), scanning through the amplitude-transition 
scale parameter space of the two e-fold transition model. When the 
amplitude of the second transition is below that of the first, i.e.\ 
$\alb$ decreases, then instability can easily arise, as seen by the 
lower shaded regions with $A_2$ below $A_1=0.1$. For strong increases in 
amplitude at early enough times, as in the upper shaded regions, this adds 
to the first transition amplitude and effectively moves the model into 
the unstable region of Fig.~\ref{fig:stabshade}. Even if the model stays 
stable, the increase in amplitude can cause $\geff$ to become stronger 
than may be observationally viable.

Finally, let us note several model independent properties. From 
Eq.~(\ref{eq:geffa2}), we see that $\geff\ge1$ for 
viable models, independent of the form adopted for $\alb(a)$. In order 
to obtain $\geff<1$, the denominator would have to pass through zero, 
which gives not only a period of divergent gravity, but instability. 
Thus, No Run Gravity is in sharp contrast to No Slip Gravity by having 
stronger gravity rather than weaker gravity than general relativity. 

In fact, we can quantify the upper limit on $\geff$ in a model independent 
manner. Rewriting the stability condition $c_s^2\ge0$ 
using Eq.~(\ref{eq:cs2}) as 
\be 
\alb\,\left(1-\frac{\alb}{2}\right)+\albp\ge \frac{3}{2}\om(a)\,\alb\ , 
\ee 
and inserting this in the denominator of $\geff$ gives 
\be 
\geff-1\le \frac{\alb(a)}{3\om(a)}\ . 
\ee 
Thus, stable models will have gravitational strength reasonably close 
to general relativity for modest $\alb$. 
[Note that this bound is necessary but not sufficient; 
at late times when $\om(a)\to0$, then $\geff-1\to\alb/(2-\alb)$.] 

The gravitational strength can be related to the gravitational 
growth index $\gamma$ \cite{groexp}, where the growth rate 
$f\equiv\om(a)^\gamma$. At early times, 
$\geff\approx 1+{\mathcal{O}}(a^{1/\tau})$ and so this corresponds to 
the $p=1/\tau>0$ case in \cite{endgro}. This gives $\gamma$ at early times 
less than the general relativity value, a characteristic of enhanced 
growth. At late times, $\geff\to\,$const, corresponding to the $p=0$ 
case (both values of $p$ are independent of the form of the transition, 
being asymptotic values). We verify numerically that, as in \cite{endgro}, 
the late time behavior of $\gamma(a)$ follows the asymptotic evolutionary 
form of general relativity.

\section{Next Generation Constraints} \label{sec:fisher} 

While the more complicated forms $\alb(a)$ can give viable results, avoiding 
regions of instability and excessively high gravity, they also significantly 
increase the number of parameters. Therefore in this section dedicated to 
observational constraints we will stay with the single transition case. 

We focus on the growth rate (times the amplitude) of the matter density 
perturbations, $\fs$, as the main observational constraint. While we 
will use the full numerical solutions of the growth equation, note that 
No Run Gravity is another example of a gravity theory successfully 
approximated by the two bin parametrization of $\gm$ \cite{mgbin}. Indeed, 
$\fs$ is accurately reconstructed over the observational redshift range 
to within 0.1\% of the numerical solution. 

The theory changes the redshift space distortion observable $\fs(a)$ 
from $\Lambda$CDM within general relativity, but such changes can be made 
by changing the background expansion as well. This is why it is useful 
to examine both the expansion history and the growth history in a 
conjoined manner. We study No Run Gravity with a \lcdm\ background and 
compare it to general relativity with other backgrounds, by means of a 
conjoined history diagram of the expansion rate $H/H_0$ and the growth 
rate $\fs$, as proposed in \cite{conjoin}. (Note \cite{conjoin} examined 
changes in both the matter density $\om$ and the dark energy equation of 
state parameter $w$, concluding that modified gravity had signatures in 
such a diagram distinct from either of them.) 

\begin{figure}[htbp!]
\includegraphics[width=\columnwidth]{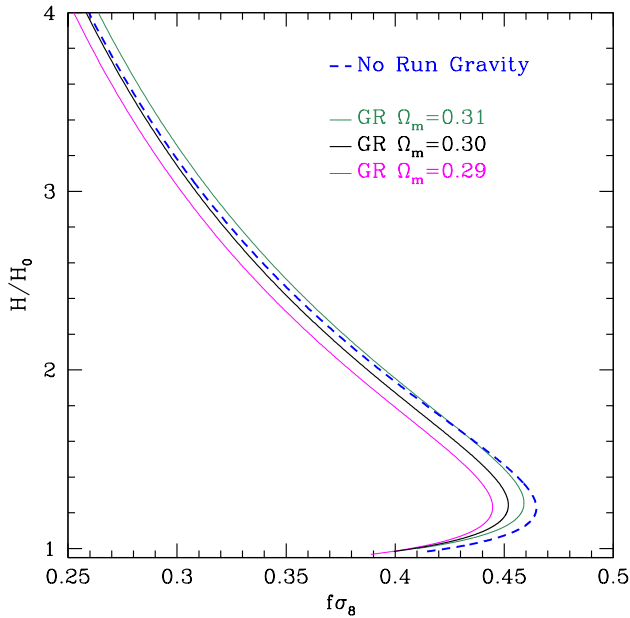}
\caption{
No Run Gravity can be distinguished from general relativity when 
considering both growth history and expansion history. Note the 
growth $\fs$ is enhanced at a given expansion rate $H$ over the 
general relativity model with the same matter density ($\om=0.3$). 
The conjoined history behavior is also distinct from a change 
in the cosmological model within general relativity. 
} 
\label{fig:conjoin} 
\end{figure}

Figure~\ref{fig:conjoin} illustrates the results, for our fiducial model 
of $\om=0.3$, $A=0.1$, $a_t=0.25$, $\tau=1/3$. Time, or scale factor, 
runs along each curve from early times at the top to today at the bottom. 
At a given expansion rate, the growth observable is greater than the general 
relativity curve with the same background. The bump, or ``nose'', of the 
modified gravity curve cuts across the general relativity models of various 
backgrounds, giving a distinctive signature.

To understand the level of constraints that may be placed on the 
modified gravity model, we consider upcoming measurements of the growth 
observable from redshift space distortion data of the Dark Energy 
Spectroscopic Instrument (DESI \cite{desi2}). This will measure $\fs$ 
over $z=0.05$--1.85 with precisions approaching percent level. We follow 
the projections given by \cite{desi1} of data precision in 18 redshift 
bins within this range, using only linear scales out to 
$k_{\rm max}=0.1\,h$/Mpc. To represent other data, such as cosmic microwave 
background measurements, we include a Gaussian prior of 0.01 on $\om$. 

\begin{figure}[htbp!]
\includegraphics[width=\columnwidth]{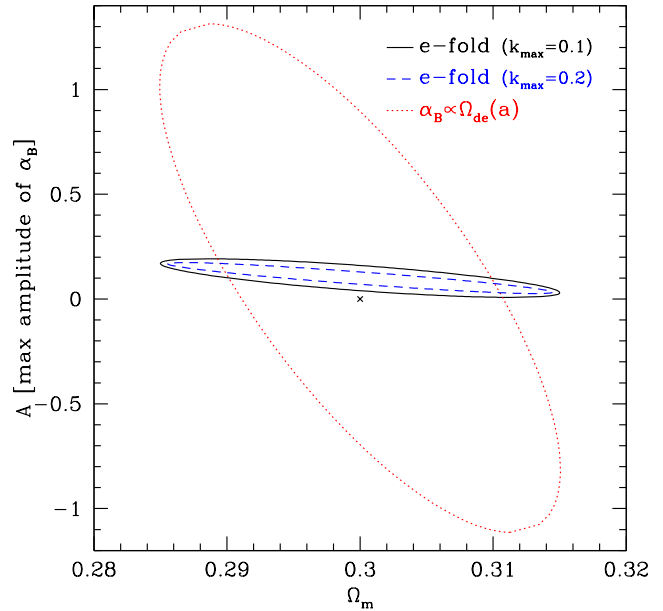}
\caption{
The joint 68\% confidence level contour for No Run Gravity is plotted 
in the maximum amplitude $A$ vs matter density $\om$ plane, for DESI 
projected measurements of the redshift space distortion observable $\fs$. 
The solid black contour uses $k_{\rm max}=0.1\,h$/Mpc; the dashed blue 
has $k_{\rm max}=0.2\,h$/Mpc. The dotted red contour shows the 
unrecommended $\alb\propto\ode(a)$ model rather than the free e-fold model. 
The x marks the general relativity case with the same expansion history. 
} 
\label{fig:elloma} 
\end{figure}

The results from a Fisher information analysis 
for the joint confidence contour of the modified gravity 
maximum amplitude and matter density are shown in Fig.~\ref{fig:elloma}. 
This uses the single e-fold transition model with a fiducial value of 
$\om=0.3$, $A=0.1$, $a_t=0.25$, and $\tau=1/3$. We fix $\tau$ to the 
fiducial value, representing the early time behavior $\alb\sim a^3$ as 
seems reasonable, and fix $a_t=0.25$, again somewhat reasonable in order 
to give a stable theory, yet one that may be connected to the onset of 
cosmic acceleration (recall $a_{t,G}=0.44$ or $z_{t,G}=1.3$). If we instead 
marginalize over these parameters then constraints are uninterestingly 
weak. We emphasize that we seek to explore indications of sensitivity, 
not carry out a detailed likelihood fit.

The constraint on the modified gravity maximum amplitude is 
$\sigma(A)=0.060$, i.e.\ a $1.7\sigma$ distinction from general relativity 
in our fiducial case. If robust density perturbation theory allows 
use of the measurements to $k_{\rm max}=0.2\,h$/Mpc, then the constraint 
tightens to $\sigma(A)=0.049$. If we had blithely extended the early time 
behavior $\alb\propto\ode(a)$ to the late time observational epoch, the 
fiducial constraints blow up to $\sigma(A)=0.8$ -- this demonstrates the 
danger of such an assumption as the transition occurs so late ($a_{t,G}=1.32$ 
or $z_{t,G}=-0.24$) that even a large modification cannot be seen in the 
data. Thus use of such a model could lead to the conclusion that general 
relativity is correct, even when the true theory is quite far from general 
relativity. 

Further data has the potential to improve our tests of modified gravity. 
Peculiar velocity surveys can probe growth at low redshifts $z\lesssim0.3$ 
\cite{adams,howlett1,howlett2} where galaxy 
clustering surveys have limited sampling and yet where the effects of 
gravitational modifications may be strongest. The surveys may use 
galaxy fundamental plane distances (e.g.\ from Taipan+WALLABY \cite{taipan} 
or an adapted DESI Bright Galaxy survey) or supernova distances (e.g.\ from 
LSST) to obtain peculiar velocities.

\section{Conclusions} \label{sec:concl} 

Upcoming data will have substantial leverage on testing gravity on 
cosmic scales. The combination of measurements in galaxy clustering, 
velocities, and lensing, CMB lensing and B-mode polarization, and 
gravitational wave standard sirens will probe the tensor sector of 
gravity, gravitational slip, and the properties of structure growth 
and light deflection. Signatures of deviations from general relativity 
in any of them would be revolutionary. 

However, structure growth and light deflection can also be affected 
by different models within general relativity. If no significant deviation 
is seen in gravitational wave propagation or gravitational slip, we may 
be tempted to say gravity is simply described by general relativity. 
Here we emphasize that this is not a necessary conclusion. We present 
No Run Gravity as a minimal modification benchmark -- one where 
gravitational waves and gravitational slip behave as in general relativity, 
but there are signatures distinct from general relativity. These 
go beyond a modification to the expansion history, i.e.\ $w(z)$, and 
beyond simple scalar fields with an extra degree of freedom in the scalar 
sound speed. 

In terms of property functions, No Run Gravity effectively has a single 
free function, the braiding $\alb(a)$. We show how this maps directly 
to the gravitational strengths for cosmic growth and light deflection, 
and that these two are both the same as each other and different from general 
relativity. The simple e-fold form is particularly attractive since then 
$\geff(a)$ is exactly a shifted, scaled form of $\alb(a)$. 

We quantified the gravitational strength and sound speed evolution, 
and analyzed the stability regions for a sound theory (indeed these 
are analytic for a particular model), as well as studying 
the gravitational growth index $\gamma$ for 
cosmic growth, related to $\geff$. 
Several relations for the stability and maximum gravitational 
strength can be written in a model independent manner -- for example for 
stable No Run Gravity $(\geff-1)\le\alb/[3\om(a)]$. 

In addition we explored how the conjoined use of expansion and growth 
data could reveal modified gravity signatures, and carried out an 
initial Fisher information sensitivity analysis for how forthcoming DESI 
data on the growth rate $\fs$ from redshift space distortion measurements 
could constrain the gravitational strength $\geff$. No Run Gravity can 
provide a benchmark for distinguishing minimal modification signatures 
from general relativity, showing the science reach of next generation 
surveys even when the more dramatic signatures may not be found.

\acknowledgments 

This work is supported in part by the Energetic Cosmos Laboratory and by 
the U.S.\ Department of Energy, Office of Science, Office of High Energy 
Physics, under Award DE-SC-0007867 and contract no.\ DE-AC02-05CH11231.


\end{document}